\documentclass{aastex631}
\usepackage{makecell}
\usepackage{amsmath}

\shorttitle{} 
\shortauthors{Shan et al.}

\graphicspath{{./}{figures/}}

\usepackage{subfigure}

\begin{document}

\title{
CAMEL. II. A 3D Coronal Mass Ejection Catalog Based on Coronal Mass Ejection Automatic Detection with Deep Learning
  }

\author[0009-0001-4778-5162]{Jiahui Shan}
\affiliation{Key Laboratory of Dark Matter and Space Astronomy, Purple Mountain Observatory, Chinese Academy of Sciences, Nanjing 210034, People’s Republic of China}
\affiliation{School of Astronomy and Space Science, University of Science and Technology of China, Hefei 230026, People’s Republic of China}

\author{Huapeng Zhang}
\affiliation{
Department of Computer Science and Technology, Nanjing University, Nanjing 210023, People’s Republic of China
}
\affiliation{
State Key Laboratory for Novel Software Technology, Nanjing University, Nanjing 210023, People’s Republic of China
}
\author[0000-0002-3032-6066]{Lei Lu}
\affiliation{Key Laboratory of Dark Matter and Space Astronomy, Purple Mountain Observatory, Chinese Academy of Sciences, Nanjing 210034, People’s Republic of China}
\affiliation{School of Astronomy and Space Science, University of Science and Technology of China, Hefei 230026, People’s Republic of China}

\author[0000-0002-9621-7321]{Yan Zhang}
\affiliation{
Department of Computer Science and Technology, Nanjing University, Nanjing 210023, People’s Republic of China
}
\affiliation{
State Key Laboratory for Novel Software Technology, Nanjing University, Nanjing 210023, People’s Republic of China
}
\altaffiliation{}
\author[0000-0003-4655-6939]{Li Feng}
\affiliation{Key Laboratory of Dark Matter and Space Astronomy, Purple Mountain Observatory, Chinese Academy of Sciences, Nanjing 210034, People’s Republic of China}
\affiliation{School of Astronomy and Space Science, University of Science and Technology of China, Hefei 230026, People’s Republic of China}

\author[0009-0000-2406-3426]{Yunyi Ge}
\affiliation{Key Laboratory of Dark Matter and Space Astronomy, Purple Mountain Observatory, Chinese Academy of Sciences, Nanjing 210034, People’s Republic of China}

\author[0000-0003-4829-9067]{Jianchao Xue}
\affiliation{Key Laboratory of Dark Matter and Space Astronomy, Purple Mountain Observatory, Chinese Academy of Sciences, Nanjing 210034, People’s Republic of China}
\affiliation{School of Astronomy and Space Science, University of Science and Technology of China, Hefei 230026, People’s Republic of China}

\author[0000-0003-2694-2875]{Shuting Li}
\affiliation{Key Laboratory of Dark Matter and Space Astronomy, Purple Mountain Observatory, Chinese Academy of Sciences, Nanjing 210034, People’s Republic of China}
\affiliation{School of Astronomy and Space Science, University of Science and Technology of China, Hefei 230026, People’s Republic of China}

\correspondingauthor{Yan Zhang, Li Feng}
\email{zhangyannju@nju.edu.cn, lfeng@pmo.ac.cn}


\received{2024 January 2}
\revised{2024 March 11}
\accepted{2024 March 19}
\published{2024 May 8}

\submitjournal{ApJS}
\begin{abstract}
Coronal mass ejections (CMEs) are major drivers of geomagnetic storms, which may cause severe space weather
effects. Automating the detection, tracking, and three-dimensional (3D) reconstruction of CMEs is important for
operational predictions of CME arrivals. The COR1 coronagraphs on board the Solar Terrestrial Relations
Observatory spacecraft have facilitated extensive polarization observations, which are very suitable for the
establishment of a 3D CME system. We have developed such a 3D system comprising four modules: classification,
segmentation, tracking, and 3D reconstructions. We generalize our previously pretrained classification model to
classify COR1 coronagraph images. Subsequently, as there are no publicly available CME segmentation data sets,
we manually annotate the structural regions of CMEs using Large Angle and Spectrometric Coronagraph C2
observations. Leveraging transformer-based models, we achieve state-of-the-art results in CME segmentation.
Furthermore, we improve the tracking algorithm to solve the difficult separation task of multiple CMEs. In the final
module, tracking results, combined with the polarization ratio technique are used to develop the first single-view
3D CME catalog without requiring manual mask annotation. Our method provides higher precision in automatic
2D CME catalog and more reliable physical parameters of CMEs, including 3D propagation direction and speed.
The aforementioned 3D CME system can be applied to any coronagraph data with the capability of polarization
measurements.
\end{abstract}
\keywords{Solar activity; Coronal mass ejections (CMEs); Astronomy data analysis}
\section{Introduction} \label{sec:intro}

Coronal mass ejections (CMEs) are the most violent and largest-scale activity in the solar system, 
which release enormous amounts of magnetized plasma. 
As commonly observed in white-light coronagraphs, 
a CME is a new, discrete, bright, white-light feature with a radially outward velocity \citep{olmedo_automatic_2008}. 
When the CME propagates toward and reaches the Earth, 
it may cause geomagnetic storms and bring catastrophic space weather. The effects of space weather can be felt in many aspects of civil life, including communications, navigation, power grids, and satellite operations. 
It is worth noting that the launch of 49 of SpaceX's Starlink satellites on 3 February 2022 failed due to geomagnetic storms \citep{dang_unveiling_2022,hapgood_spacexsailing_2022}.
Therefore, CME detection, tracking and three-dimensional (3D) reconstruction are important not only for physical studies but also for the monitoring of space weather environment.

Thanks to the decades of observation by the Solar and Heliospheric Observatory (SOHO), 
especially those by the Large Angle and Spectrometric Coronagraph Experiment \citep[LASCO, ][]{brueckner_large_1995},  
we have been able to study CMEs and their relationship to the solar cycle. 
To further understand CMEs, especially its 3D evolution, 
the twin Solar Terrestrial Relations Observatory (STEREO) spacecrafts are equipped with an instrument suit called the Sun Earth Connection Coronal and Heliospheric Investigation \citep[SECCHI,][]{howard_sun_2008}.
With the accumulation of coronal images, it becomes increasingly important to identify and catalog CMEs.

As a widely adopted catalog, coordinated data analysis workshops \citep[CDAW, ][]{yashiro_catalog_2004} is maintained manually, 
which is time-consuming and may be subjective, particularly during times of high solar activity. 
The shortcomings of manual catalogs have led to the development of automated methods for detecting and tracking CMEs.
Some methods deployed their own CME catalogs utilizing LASCO and/or SECCHI coronagraph images, 
such as the first automatic detection method named the Computer Aided CME Tracking Catalog \citep[CACTus, ][]{robbrecht_automated_2004,robbrecht_automated_2009};
the solar eruptive event detection system \citep[SEEDS,][]{olmedo_automatic_2008};
the automatic recognition of transient events and marseille inventory from synoptic maps (ARTEMIS) I and II \citep{boursier_artemis_2009,floyd_artemis_2013};
Coronal Image Processing \citep[CORIMP,][]{byrne_automatic_2012}. 
By employing CORonal SEgmentation Technique (CORSET) method \citep{goussies_tracking_2010} on for STEREO/COR 2 observations, 
a dual-viewpoint CME catalog was established \citep{vourlidas_multi-viewpoint_2017}. 

\cite{qiang_cme_2019} presented an adaptive background learning method, 
where CMEs are moving foreground objects in the background. 
\cite{zhang_detection_2017} employed a machine learning approach known as Extreme Learning Machine (ELM) to identify suspicious CME regions based on the brightness and texture features within the images. 
These automatic detection methods primarily rely on texture features, gray features, optical flow methods, and traditional machine learning techniques.
As simple thresholds or artificially selected features are only part of CME features, recognition results are not accurate enough.

Traditional machine learning has also been applied to the prediction of the CME arrival to the Earth. \cite{liu_new_2018} utilized Support Vector Machine (SVM) to analyze various parameters of CMEs for predicting the transit time of CMEs from the Sun to the Earth. \cite{yang_prediction_CMEs} employed the ensemble learning method XGBoost and introduced two effective feature importance ranking methods.

Compared with traditional machine learning techniques, deep learning has a more powerful feature extraction capability.
\cite{wang_new_2019} introduced convolutional neural network (CNN) to recognize CME and proposed a method called CME automatic detection and tracking with machine learning (CAMEL).
As supervised learning requires a large number of labels, they simply used the CDAW catalog. 
However, some CMEs are missed or mislabeled in the CDAW catalog, particularly narrow CMEs that occurred during the solar maximum. 
Thus, \cite{jia-hui_automatic_2020} manually relabeled coronagraph running-difference (RDF) images and made a dataset for classification. 
The Visual Geometry Group (VGG) 16 model \citep{simonyan_very_2015} was applied to classify images with or without CMEs. 
\cite{alshehhi_detection_2021} proposed an unsupervised method to detect CME, where pre-trained VGG 16 model weights were used to extract features. 
Following principal component analysis (PCA) to reduce dimensionality, binary classification was performed for each pixel by K-means clustering. 
These methods, however, result in the detected coarse CME structures.

In recent years, artificial intelligence and deep learning techniques have been flourishing. 
In the field of computer vision, deep learning has achieved excellent performance in image classification, object detection, image segmentation, image super-resolution, etc.  
Image classification and semantic segmentation can be both treated as classification tasks.
The CME image classification only predicts a category for the entire image,
whereas the CME semantic segmentation involves predicting a category for each pixel. 
In 2015, CNNs demonstrated significant performance in semantic segmentation tasks. 
Described as a seminal work, Fully Convolutional Network \citep[FCN,][]{long_fully_2015} solves the segmentation problem by removing the fully connected layers. 
On the basis of its encoder-decoder architecture, CNN-based semantic segmentation models have been proposed from different aspects to refine the coarse prediction of FCNs. 
To enlarge the receptive field, Dilation \citep{chen_semantic_2016} and DeepLab \citep{yu_multi-scale_2016} adopted dilated convolution. 
To obtain global contextual information differing across sub-regions, 
Pyramid Scene Parsing Network \citep[PSPNet,][]{zhao_pyramid_2017} utilizes large kernel pooling layers.  
Moreover, \cite{fu_dual_2018} and \cite{huang_ccnet_2018} designed different attention modules combined with CNNs.

In recent research, vision transformers have demonstrated superior performance for computer vision tasks. 
Originally, transformer \citep{vaswani_attention_2017} was designed for natural language processing (NLP) tasks. 
The significant success that transformers have achieved in NLP \citep{devlin_bert_2018,floridi_gpt-3_2020} has impelled researchers to apply them to computer vision (CV). 
In 2020, \cite{2020arXiv201011929D} first proposed a pure transformer to replace CNNs and achieved state-of-the-art performance on image classification tasks. 
Subsequently, a series of vision transformer models have been developed to improve various computer vision problems \citep{2020arXiv201212556H,jamil_comprehensive_2022}. 
In image segmentation, the SEgmentation TRansformer (SETR) proposed by \cite{zheng_rethinking_2021} demonstrates the feasibility of a pure transformer in this task. 
To generate multiscale feature maps for dense prediction, \cite{wang_pyramid_2021} proposed a pyramid vision transformer (PVT) with pyramid structures. 
\cite{xie_segformer_2021} redesigned the semantic segmentation framework and proposed SegFormer, which combines a hierarchical transformer encoder with a lightweight multilayer perception decoder for generating multi-scale features. 
In comparison with CNNs, the simple, efficient yet powerful SegFormer reached the state-of-the-art segmentation performance. 

In summary, deep learning is still in its infancy in terms of CME detection, primarily due to the lack of publicly available labeled datasets.
Since image labels are readily available from open CME catalogs, most machine-learning methods for detecting CMEs are still in the stage of image classification. 
It is worth noting that the study of the segmentation of CME structural regions is also important.

Studying the 3D structure of CMEs provides insight into the possible relationships between the initial evolution of CMEs, 
their subsequent propagation, and their effects on the Earth.
\cite{mierla_3d_2009,mierla_3-d_2010,mierla_low_2011} and \cite{feng_soph_2013} demonstrated a series of representative techniques for determining the location and the geometry of CMEs in the 3D space.
They compared the results obtained by applying these techniques to coronagraph observations from SECCHI and LASCO.
Among these techniques, the polarization ratio technique, developed by \cite{moran_three-dimensional_2004} based on Thomson-scattering theory, could locate the center of mass along the line of sight (LOS) for each pixel within a CME from only one perspective \citep{dere_three-dimensional_2005,moran_three-dimensional_2010,susino_three-dimensional_2014}. 
However, these methods require manually masking out the CME region first.
Currently, there are no automated methods for 3D reconstruction of CMEs, nor are there any publicly available 3D CME catalogs.

We are accumulating a large amount of coronagraph polarization data observed by, e.g., COR1 in the instrument suit SECCHI \citep{howard_sun_2008} aboard STEREO, the Multi Element Telescope for Imaging and Spectroscopy \citep[METIS,][]{antonucci_metis_2020} aboard Solar Orbiter, Lyman-alpha Solar Telescope \citep[LST,][]{li_lyman-alpha_2019,chen_lyman-alpha_2019,feng_lyman-alpha_2019} aboard Advanced Spaced-based Solar Observatory, Association of Spacecraft for Polarimetric and Imaging Investigation of the Corona of the Sun \citep[ASPIICS,][]{renotte_ASPIICS_2015} aboard Proba-3. An automatic 3D catalog is vital for the operational prediction of the CME arrival at Earth and for statistical studies of CME physics. 
Therefore, We have developed an automatic system for detecting, tracking, and 3D reconstructing CMEs. Additionally, we have published both a two-dimensional (2D) catalog and a 3D CME catalog online. (Please visit the catalog websites written at https://github.com/h1astro/CAMEL-II).
The automatic recognition of COR1 is more challenging than LASCO C2, mainly due to the increased difficulty in suppressing stray light from the inner occulting coronagraph, which introduces more non-coronal structural interference in the images. Our previous work in 2019 \citep{wang_new_2019} was primarily based on LASCO C2, and now, this study is focused on COR1.
The Transformer-based segmentation model is applied to CME segmentation for the first time and achieves state-of-the-art performance. Furthermore, we validate the generalization capability of the trained classification and segmentation models for COR1 data aboard the STEREO Ahead spacecraft (hereafter COR1-A) which were originally developed for LASCO C2 data. An improved tracking algorithm that yields more accurate and reliable structural evolution of CMEs shows the capability of separating successive CMEs. It is a difficult task in cases of the occurrence of multiple CMEs. Besides these improvements for 2D images, to derive true 3D propagation direction and speed of CMEs, we also automate their 3D reconstructions using the polarization ration method for each pixel in the CME mask defined by the convex hull of the segmented CME area. To our knowledge, it is the first automated 3D CME catalog.

This paper is structured as follows:
In Section \ref{sec:data}, we provide a description of the two selected coronagraphic observations and outline the preprocessing operations conducted on the data.
Section \ref{sec:Methodology} is divided into two main parts. One part presents the methodology employed for the COR1-A data, which encompasses image classification (Section \ref{subsubsec:classification_cor1_a}), image segmentation (Section \ref{subsubsec:segmentation_cor1_a}), improved tracking algorithms (Section \ref{subsubsec:tracking_cor1_a}), and the automatic reconstruction of CMEs in three-dimensional space (Section \ref{subsubsec:PR_cor1_a}).
The other part introduces the labeled LASCO C2 segmentation dataset (Section \ref{subsubsec:segmentation_dataset}) and provides details about the architecture of the Transformer-based segmentation model SegFormer (Section \ref{subsubsec:segformer_model}).
Section \ref{sec:Results} presents an analysis of the experimental results, including segmentation, tracking, and 3D reconstructions.
Finally, we conclude the paper in Section \ref{sec:summary}.

\section{Observational Data and Image Preprocessing} \label{sec:data}
The white light coronagraph observations used in this paper were obtained from LASCO C2 and SECCHI/COR1-A. 
LASCO C2 has a field of view (FOV) of 2.2 to 6~$\mathrm{R_\odot}$. COR1 has a FOV of 1.3 to 4 $\mathrm{R_\odot}$ with additional capability of linear polarization measurements.

LASCO C2 data is used to train the segmentation network.
We aim to verify the generalization performance of the segmentation model trained on LASCO C2 data when applied to other coronagraph observations, e.g., COR1-A.
Using lasco\_readfits.pro and reduce\_level.pro from the SolarSoftware, 
the level 0.5 FITS files were converted into level 1 data. 
The processing includes corrections for dark current, flat field, stray light, distortion, vignetting, photometry, time, and position. 
Then, the resolution of an image was downscaled from $1024 \times 1024$ to $512 \times 512$. 
A $3 \times 3$ mean filter is used to suppress sharp noise features in the downsampled images. 
By subtracting the previous frame from the current image, the running-difference image $P_\text{{rdf}}$ is produced, which is used as input to the segmentation network. 

Next, COR1-A running-difference images are tested by a trained segmentation model.
Level 0.5 FITS files were downloaded. 
Similar to the LASCO C2 preprocessing steps, 
the images were calibrated using secchi\_prep.pro to process the Level 0.5 to Level 1 data. 
Then, a $2 \times 2$ median filtering operation was used. Based on the observation log of COR1-A, we deleted the data generated during the calibration roll and momentum dump since these could impact our tracking results.

\section{Methodology} \label{sec:Methodology}
\begin{figure}[ht!] 
    \plotone{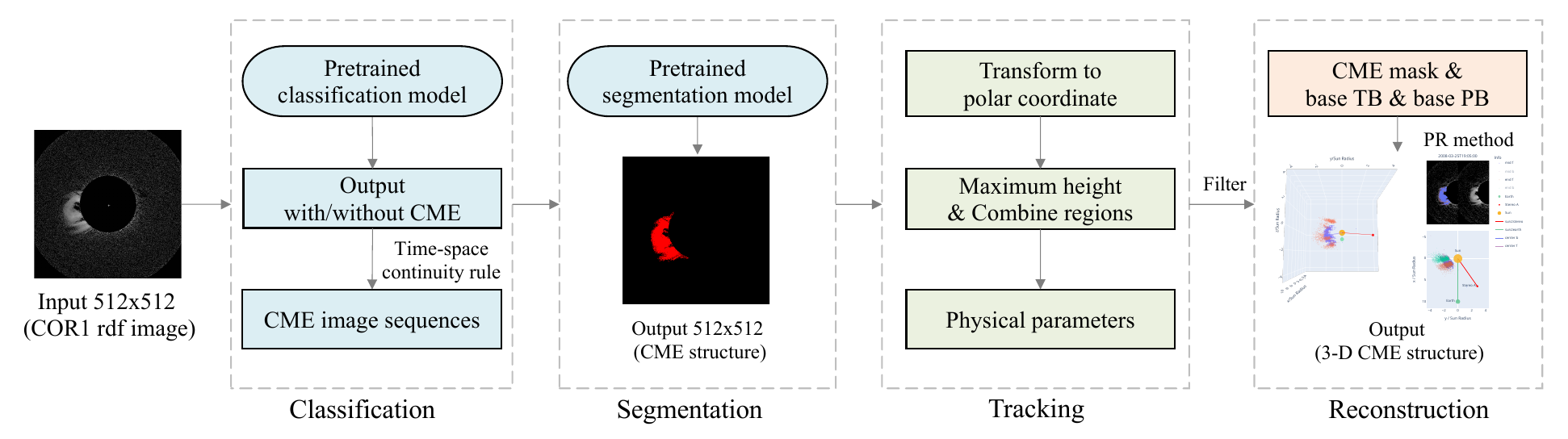}
    \caption{Flowchart for automatically reconstructing three-dimensional CMEs\label{fig:flowchart}}
\end{figure}

We propose a method that can automatically derive CME image sequences, create masks of CME regions, and reconstruct 3D CMEs. As shown in Figure \ref{fig:flowchart}, the pipeline consists of four major modules: classification, segmentation, tracking, and 3D reconstructions.
In the classification module, after obtaining labels with or without CME through our pre-trained VGG model, CME image sequences are divided according to the time-space continuity rule. 
Similarly, for the CME segmentation module, 
we simply feed CME image sequences into our pre-trained segmentation model without fine-tuning and obtain pixel-level CME regions.
To train the segmentation model, we manually annotated the LASCO C2 segmentation dataset.
The third module involves the tracking algorithm to get more complete CME image sequences.
Our final module focuses on the 3D reconstruction of CMEs using the polarization ratio (PR) technique.

\subsection{CME Image Classification based on Pre-trained Model} \label{subsubsec:classification_cor1_a}
Our pre-trained VGG model \citep{jia-hui_automatic_2020} was employed to classify the input running-difference image for the presence of CMEs. 
This model yields a binary label: 1 is the presence of CMEs in the image, while 0 denotes their absence. 
The classification dataset was derived from the LASCO C2 dataset.
Because some images containing CME structures are not recorded in the CDAW CME catalog, we collected and labeled 4483 images manually for training.
Considering the similarity of running-difference images between LASCO C2 and SECCHI COR1-A, we generalize the classification and segmentation models trained on LASCO C2 to COR1-A.

Since CMEs evolve over time, we can use the labels obtained from the pre-trained classification model to divide these images into CME image sequences, which represent preliminary CME events. Note that at this step, a CME image sequence may contain one or more individual CME events. Each image sequence will be divided into individual CMEs in the tracking module in Section~\ref{subsubsec:tracking_cor1_a}. We adhered to the spatiotemporal continuity rules described in our previous work to obtain preliminary CME events.
The first requirement is that a CME image sequence may contain an image labeled 0 (without CMEs) but cannot contain more than two consecutive images labeled 0. 
One image labeled 0 might be due to misclassification. However, two consecutive misclassifications are unlikely and exceed the model tolerance.
Lower limits are defined for the total time and number of images in the sequence to determine whether to discard this image sequence.
Lastly, we calculate the time difference among image sequences, 
and set a time threshold of one hour to decide whether to merge the retained image sequence into either the previous or next image sequence, or keep the retained image sequence isolated. 

Furthermore, we made two additional rules to acquire better results, in particular, to improve the accuracy of the CME start time.
In each preliminary CME image sequence, the previous image before the first image would be added if the time interval between the two images does not exceed one hour. 
In the segmentation pipeline, if the preceding image in the current sequence is classified as non-CME by the classification model but is segmented by the segmentation model to reveal CME regions, we include this preceding image in the CME image sequence. Because in the initial stages of CME evolution, where its shape is very small, it is prone to be classified as non-CME by the classification model. 
Due to lack of data or invalid images, e.g., during the calibration rolls, if a temporal gap exceeds 90 minutes between two consecutive images, we consider it unlikely to be the same CME event and split the CME image sequence into two.

\subsection{CME Image Segmentation based on Transformer-based Model} \label{subsec:Segmentation}
We established our own CME segmentation dataset based on LASCO C2 data and trained two segmentation models: a CNN-based model PSPNet and a Transformer-based model SegFormer.
The PSPNet is founded on the concept of integrating a pyramid pooling module, which aims to capture contextual information across varying scales. 
However, the Transformer-based model has a larger effective receptive field.
Since SegFormer yielded superior results (see Table \ref{table:segmentation_compare} with more details in Section~4.2), we will focus on detailing this model.
Then, the trained SegFormer model was employed for COR1-A coronagraph observation.
Semantic segmentation aims to classify each pixel in the image but does not distinguish between objects in the same category. 
So after the segmentation pipeline, multiple CME events cannot be identified simultaneously.
In the tracking section \ref{subsubsec:tracking_cor1_a}, 
we will provide specifics on separating two CMEs in an image if the angular width between them exceeds a certain threshold. 

\subsubsection{A New Segmentation Data Set} \label{subsubsec:segmentation_dataset}
This paper tries to derive more accurate CME regions by segmenting running-difference images using a transformer-based model. It is necessary to collect a semantic segmentation data set before using the segmentation model to learn the CME structures. However, building a CME segmentation dataset is significantly more complicated than a classification dataset for the following reasons.
First, we need to annotate the specific details of the CME structure. We cannot simply judge whether there is a CME, and neither can we use a rectangle to approximate the position of CMEs.
Second, the CME structure edge is often obscure, making it more challenging to annotate than the usual segmentation data. Additionally, when annotating CMEs, we also marked shock waves. 
Third, it is time-consuming to search for and annotate hundreds of images with varying angular widths.

We defined four types of CMEs with respect to their angular widths: narrow (width$\leq$30°), limb (30°$\textless$width$\leq$ 120°), 
and partial halo (120°$\textless$width$\textless$ 360°), full halo (width = 360°) which is similar to the CDAW catalog \citep{yashiro_comparison_2008,alshehhi_detection_2021}.
To cover these four types of CMEs, we gathered a total of 375 LASCO C2 images.
To span the observations at different levels of solar activity and take into account the instrument degradation, observations taken in April 2011, February, March,
September, November 2012, and January 2017 were chosen.
The proportion of narrow, limb, partial halo, and full halo CMEs in the dataset is approximately 6:8:3:1. Then, CME regions are manually marked in running-difference images to create a segmentation dataset.
The training and validation sets are divided randomly by 4:1.


\subsubsection{Architecture of SegFormer Model} \label{subsubsec:segformer_model}
\begin{figure}[ht!]
  \plotone{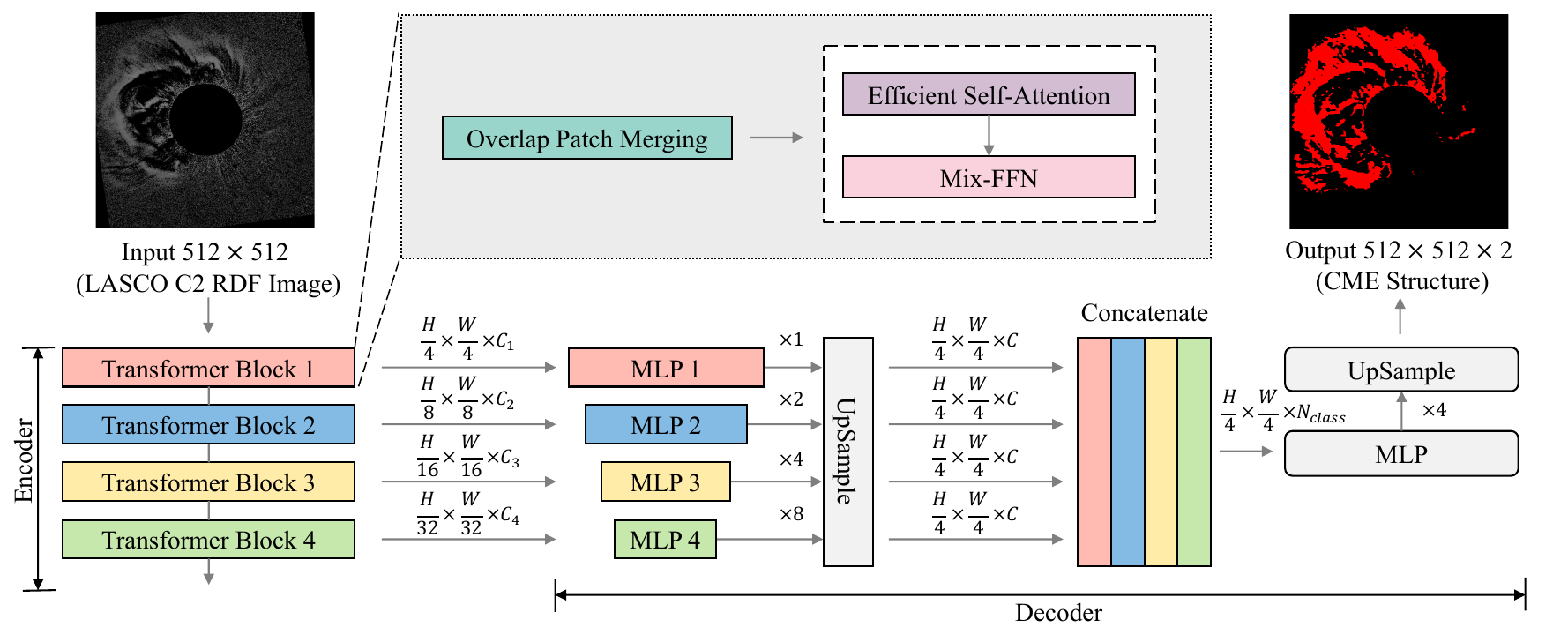}
  \caption{The SegFormer segmentation model uses running-difference images as input and predicts the structure of CMEs as output.
SegFormer's architecture primarily comprises two components: an encoder and a decoder.
The Mix Transformer encoder consists of four different transformer blocks.
There are three modules within each transformer block: overlap patch merging, efficient self-attention, and mix feed-forward network (Mix-FFN).
$H$ represents the height of the input image, $W$ denotes its width, and $C_i (i \in {1, 2, 3, 4})$ is the number of channels in the output from each transformer block's final dimension.
The decoder consists of four multilayer perceptron (MLP) layers and predicts the semantic segmentation mask. Where $C$ is the number of channels, $\mathrm{N_{class}}$ is the number of categories.}\label{fig:model_segformer}

\end{figure}

SegFormer is an encoder-decoder semantic segmentation model based on the Transformer architecture, 
which has demonstrated superior performance compared to CNNs. 
The performance of the vision transformer (ViT) on semantic segmentation of CME running-difference images is yet to be determined.
As depicted in Figure \ref{fig:model_segformer}, this model is applied to the segmentation of CMEs.

\emph{SegFormer Encoder.}
The Mix Transformer (MiT) encoder consists of four different transformer blocks, to extract coarse features with high resolution and fine-grained features with low resolution.
There are three modules within each transformer block: 
overlap patch merging, efficient self-attention (ESA), and mix feed-forward network (Mix-FFN).

The overlap patch merging makes use of a convolution layer to reduce the resolution of the feature map. 
Additionally, adjacent patches partially overlap when the patch size, stripe, and padding size are adjusted. 
It avoids the problem caused by ViT's patch merging method, 
which loses the local continuity around those patches.
For instance, a running difference image may be sliced into several pieces and scattered within adjacent image patches.

Self-attention \citep{vaswani_attention_2017} is the core concept of the Transformer architecture. It addresses the issue of non-parallelizable computation in recurrent neural networks (RNNs), focusing on the key features and easily capturing long-range dependencies, thus providing a larger effective receptive field.
Traditional self-attention has a query $\it{Q}$, a key $\it K$, and a value $\it V$ as inputs. The dimensions $d_{\it k}$ of $\it{Q}$, $\it{K}$, and $\it{V}$ are all $N \times C$, where $ N = H \times W$ is the sequence length, $H$ is the input image's height, and $W$ is its width.
The attention operation is presented in Equation (\ref{equation:attention}),
where a softmax function is applied to obtain the weights on the values. 
Unlike traditional self-attention, ESA's $K$ and $V$ have smaller sequence lengths $\overline{N}=N/R$, implemented via a convolution layer, where $R$ is the reduction ratio. 
ESA uses sequence reduction \citep{wang_pyramid_2021} to reduce computational complexity from $O(N^2)$ to $O(\frac{N^2}{R})$, 
which is especially advantageous for processing large-resolution images.

\begin{equation}
  \rm{Attention}\it(Q,K,V) = \rm{softmax}(\it \frac{ QK^T }{ \sqrt{d_{\it k}} } )V 
  \label{equation:attention}
\end{equation}

Encoding positional information for different resolutions requires interpolation. 
Thus, Mix-FFN replaces positional encoding (PE) in ViT with a $3\times3$ convolution layer. 
By implicitly learning patch positional information, Mix-FFN can avoid degradation of accuracy.
Here is the formula for implementing Mix-FFN:

\begin{equation}
  \rm X_{\it  out} = \rm{MLP}(\rm{GELU}({Conv}_{3\times 3} ({MLP}( X_{\it ESA} )))) + X_{\it ESA}
\end{equation}

where $\rm X_{\it ESA}$ represents the feature from the ESA module, MLP refers to multilayer perceptron comprising fully connected neurons with at least three layers and employing a nonlinear kind of activation function, GELU is Gaussian error linear unit.


\emph{SegFormer Decoder.}
In comparison to traditional CNN encoders, SegFormer's hierarchical transformer encoder has a larger effective receptive field.
Accordingly, SegFormer employs a lightweight decoder consisting of only multilayer perceptron layers and reduces the amount of computation.

First, the decoder outputs four multi-level feature maps with different channel dimensions. 
Each feature map $X_i (i \in \{1,2,3,4\})$ is unified to the same dimension C through a linear layer, respectively.
The second step is to upsample all of these feature maps to $\frac{H}{4}\times \frac{W}{4}\times C$ using bilinear interpolation.
Third, the concatenated features are fused together using a linear layer. 
Fourth, another linear layer is utilized to predict the mask $M_p$ with $\frac{H}{4}\times \frac{W}{4}\times N_{class}$ resolution for the fused features, where \textbf{$\mathrm{N_{class}=2}$} represents the number of categories. 
Lastly, in our pixel-level CME segmentation, each pixel is classified into two categories, either with or without CME.
A $512\times 512\times 2$ output mask $M$ is generated after upsampling, matching the input image's length and width.
An outline of the SegFormer decoder is shown in Equation (\ref{equation:decoder}).

\begin{equation} 
\begin{aligned}
X_i & = {\rm MLP}_{(C_i,C)}(X_i), {\rm where}\, i \,{\rm \in \{1,2,3,4\}} \\ 
X_i & = {\rm Upsample}_{\frac{H}{4} \times \frac{W}{4}}(X_i) \\
X & = {\rm MLP}_{(4C,C)}({\rm Concat}([X_1,X_2,X_3,X_4])) \\
M_p & = {\rm MLP}_{(C,N_{class})}(X) \\
M & = {\rm Upsample}_{H \times W}(M_p)
\label{equation:decoder}
\end{aligned}
\end{equation}

\subsubsection{COR1-A CME Image Segmentation based on Pre-trained Model} \label{subsubsec:segmentation_cor1_a}
Similarly to the CME classification, 
the SegFormer model we pre-trained on the LASCO C2 dataset is applied to the COR1-A data.
By feeding COR1-A CME image sequences into the model, CME regions can be segmented without fine-tuning.
The majority of images without CME structure are deleted after the classification module. 
Still, there are some images that are incorrectly labeled as CME. 
The purpose of the segmentation module is to extract the regions containing CME structure as accurately as possible and remove those incorrectly labeled images.

\subsection{Tracking CME Events} \label{subsubsec:tracking_cor1_a}

We aim to track the structural evolution of CME events over time through image sequences. 
It should be noted that an image sequence derived from the classification module is coarse and may have multiple CME events. 
Additionally, a segmented image may include at least one CME or contain significant noise or even only noise. 
We hope to have a tracking algorithm to solve these problems. 
When we used the previous tracking algorithm \citep{wang_new_2019} to establish a one-year catalog, 
tracking results were not satisfactory even after adjusting relative parameters about COR1-A's FOV.  
Accordingly, we propose several improvements in the tracking algorithm.

\begin{figure}[htbp]
  \centering
  \subfigure[]{%
  \includegraphics[width=4cm]{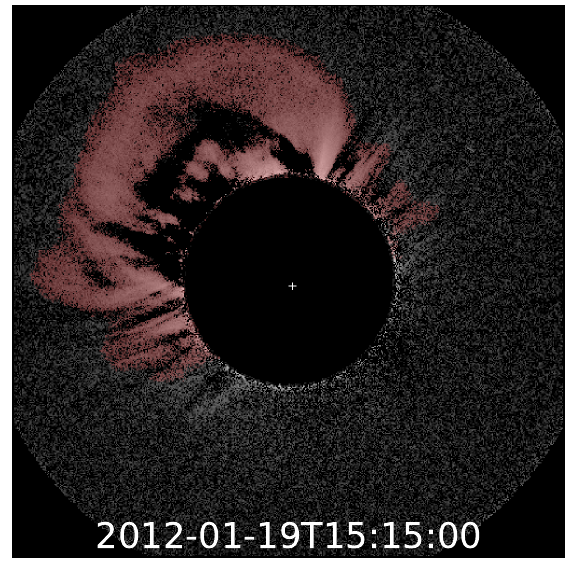}
  \label{fig:track_rule_ori}}
  \quad
  \subfigure[]{%
  \includegraphics[width=4.25cm]{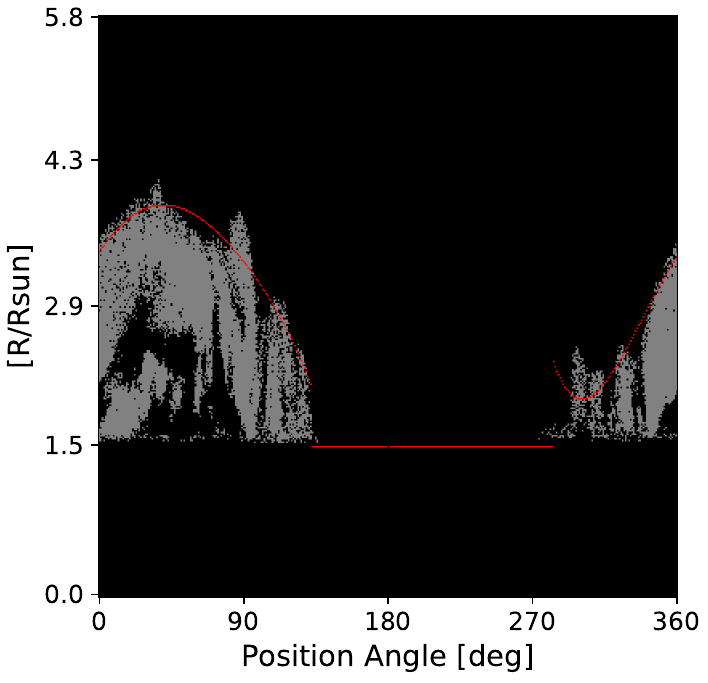}
  \label{fig:track_rule_fit_height}}
  \quad
  \subfigure[]{%
  \includegraphics[width=4.4cm]{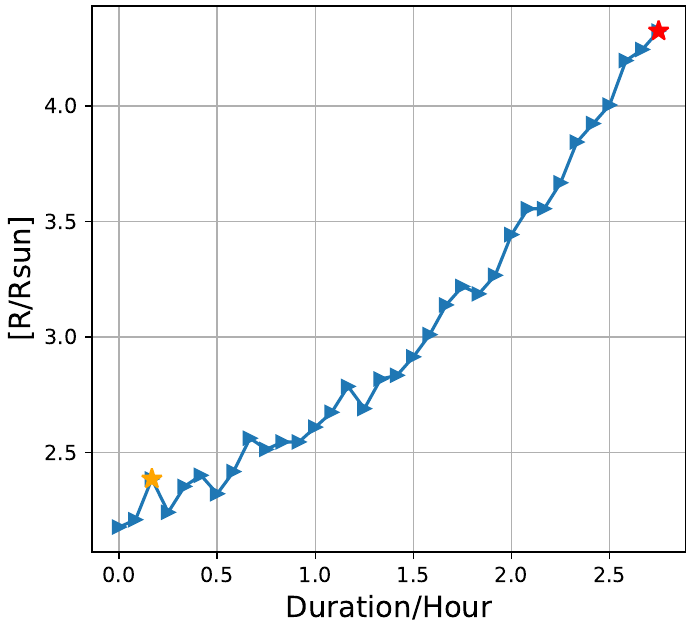}
  \label{fig:track_rule_speed_height}}
  
  \subfigure[]{%
  \includegraphics[width=5.9cm]{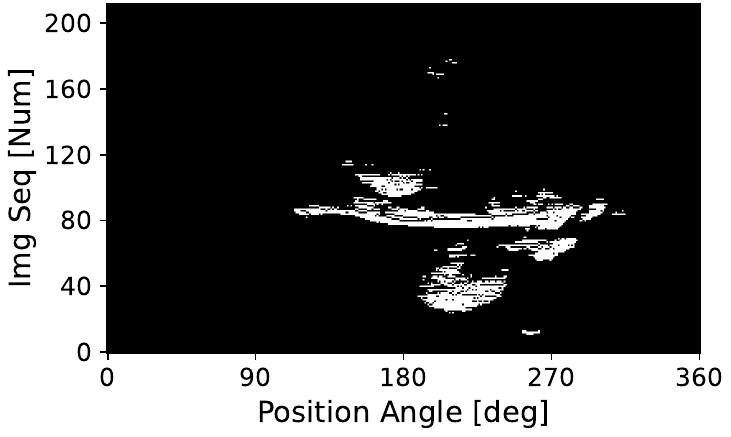}
  \label{fig:track_rule_sequence}}
  \quad    
  \subfigure[]{%
  \includegraphics[width=5.9cm]{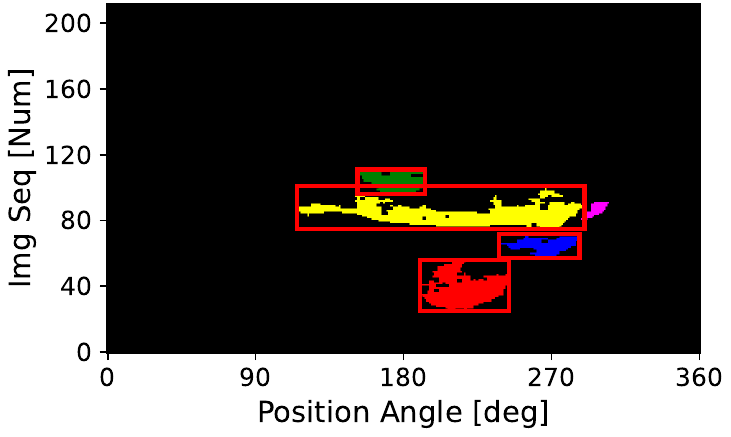}
  \label{fig:track_rule_sequence_split}}

  \subfigure[]{%
  \includegraphics[width=14.0cm]{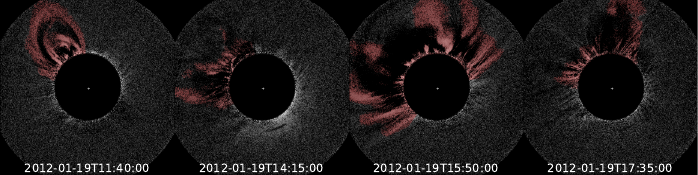}
  \label{fig:track_rule_results}}

  \caption{
  Demonstration of the CME tracking module. 
(a) The red part in the running-difference image is the segmented pixel-level CME area. 
(b) Transformation of the red segmented CME regions to the gray regions. The red line represents 1.5$R_\odot$, the minimum height we defined. The red curve is obtained by fitting the maximum height at each position angle (PA). 
(c) CME height-time diagram at the PA with median velocity. For each blue triangle, the abscissa represents the time-lapse of each image to the first image in the CME image sequence, while the ordinate represents the CME height at the PA.
The yellow star denotes the position where an increase in height is strictly required. The red star indicates the fulfillment of our current rules, and all height points during this period are used for fitting the velocity.
(d) The stack of CME detection, plotted in time, i.e., image number, against PA measured counterclockwise from solar north.
Each row corresponds to an image. White dots within each row indicate outward propagation at these PAs in the subsequent frame.
(e) Each red bounding box represents a divided image sequence with different colors. Small gaps are eliminated using a closing morphological operator. 
(f) Four distinct CME events were successfully separated from an image sequence.
The four extracted frames of CME events are taken from the ending boundaries of the bounding boxes. These masks were overlaid onto the corresponding running-difference images.
  \label{fig:post_processing_rule}
  }
\end{figure}

Before tracking, 
it is necessary to post-process the segmentation result to improve tracking precision. 
There are three major steps in our postprocessing: 
removing invalid or error images, separating sequences based on time continuity, 
and dividing sequences based on space-time continuity. 
\begin{itemize}
\item[$\bullet$] \emph{Removing invalid or error images.}
Each binary image of the CME borders is obtained using convex hull morphological operations. 
These binary images with fewer than $100$ white pixels are considered to be dominated by noise and are deleted.
Polar transformation is applied, starting from the North of the Sun going counterclockwise.
Each segmented image is transformed with a $\theta \times r$ resolution to count the CME structure at each position angle (PA),
where $\theta$ is the polar angle (i.e., PA) and $r$ is the radial distance measured from the limb.
Similarly, the segmented images are discarded if the angles are less than five degrees.

\item[$\bullet$] \emph{Separating sequences based on time continuity.} 
In COR1 FOV, CMEs with speed of $~\mathrm{250~kms^{-1}}$ may be regarded as relatively slow CMEs. This speed corresponds to a duration of two hours in COR1 FOV which covers 24 frames considering COR1's cadence of 5 minutes. A CME image sequence with more than 24 frames may suggest different CME events. For such an image sequence, the time difference $T_{\text{diff}}$ between two consecutive images is calculated. If $\mathrm{T_{\text{diff}} > 2~hours}$, we divide the image sequence into multiple sub-image sequences. On the other hand, CME image sequences containing fewer than $3$ images are excluded. The number of 24 and threshold of $T_{\text{diff}}$ can also be reduced to distinguish two CMEs that occur with a very short time lapse.
This step serves as a coarse separation of a CME sequence.

\item[$\bullet$] \emph{Dividing sequences based on space-time continuity.}
According to the classification module and the previous tracking algorithm \citep{wang_new_2019}, 
it is difficult to distinguish multiple CMEs whose PAs overlap but erupt at different times in an image sequence.
The CMEs that break out late are often overlooked, 
particularly in years around solar activity maximum, 
when multiple CMEs sometimes occur closely in time.
To overcome this challenge, we propose a method that combines the temporal and structural characteristics of CMEs to temporally partition the sub-image sequences. 
CME image sequences with more than 24 images are transformed into a $[\theta, r]$ polar coordinate system. In the tracking stage, the input is the CME region segmented from each image in the image sequence.
Figure \ref{fig:post_processing_rule} provides an illustrative example, 
depicting the transformation of the red segmented CME regions in Figure \ref{fig:track_rule_ori} to the gray regions in Figure \ref{fig:track_rule_fit_height}. 
The transformed images have a resolution of $360 \times 360$ and a radial field of view ranging from $1.3$ to $4 R_\odot$. 
In the transformed image $n_i$ of the CME image sequence, $H[n_i, \theta_j]$ represents the maximum height of the CME mask at the PA $\theta_j$. 
A CME, commonly observed in white-light coronagraphs, manifests as a discrete, bright, white-light feature with a radially outward velocity. 
Figure \ref{fig:track_rule_sequence} demonstrates the identification of propagating angles of the CME structure by comparing the height differences between contiguous images. 
The presence of white pixels in a row indicates the recorded angles of outward propagation. 
To address small gaps, a closing morphological operator is applied in Figure \ref{fig:track_rule_sequence_split}. 
Subsequently, a label morphological operator is employed to identify connected regions with different colors labeled and remove small objects. 
The boundaries of the same color regions are outlined by bounding boxes. Bounding boxes with significant height overlap are merged.
The time corresponding to the first image of the bounding box is used as the start time $T_{\text{start}}$ of the new sub-image sequence.
The image corresponding to the maximum angle signifies the moment when the CME angular width reaches its peak, denoted as $T_{\text{peak}}$.
As seen in Figure \ref{fig:track_rule_results},
our improved tracking algorithm can successfully solve the challenging task of distinguishing multiple CMEs in an image sequence. 

\end{itemize}

After the post-processing stage, we proceed to track the updated image sequence to capture the structural evolution of each CME and extract its associated key physical parameters. 
We have made substantial improvements to the tracking algorithm initially proposed by \cite{wang_new_2019},
which follows similar tracking rules as described in \cite{olmedo_automatic_2008}.
To facilitate the tracking process, the image sequences are converted to the polar coordinate system, 
and the maximum height at each PA is calculated. These discrete PAs are then merged into one or more angular widths. 
We define $N_{fov}^{\theta}$ as the number of frames in which the maximum height of each PA in an image sequence exceeds the minimum height threshold. 
The previous rule for determining the minimum height threshold was based on half of the sum of the maximum and minimum fields of view. 
However, this led to the omission of many CME events in COR1 images. After careful adjustments, we have set the minimum height threshold to $2.3 R_\odot$.
To identify the presence of CMEs at specific PAs, 
we assign a value of 1 to $L_{\theta_j}$ if $N_{fov}^{\theta_j}$ (
$L_{\theta}$ is a $1 \times 360$ vector, the number of $N_{fov}^{\theta}$ at PA $\theta_j$) is at least two; otherwise, it is marked as 0.

The change in radial heights along a PA is sometimes non-incremental due to various factors such as the evolution of CME structure, noise, and segmentation errors. 
As indicated by the red line in Figure \ref{fig:track_rule_fit_height}, we fit a height curve to the PA-height diagram.
In addition, the speed calculation rule has been modified so that radial heights in the time series do not need to increase continuously, 
while some radial heights are allowed to decrease slightly.
In this way, a linear fit is done for the points that meet the rules to calculate the speed more accurately as depicted in Figure \ref{fig:track_rule_speed_height}.
By calculating the velocity distribution at PAs around the central PA, 
we choose the median and maximum values as the median and maximum velocity of the CME, respectively.
The end time of the CME is difficult to determine, 
we choose the moment when a CME reaching the maximum height or running out of the COR1 FOV as the end time.

In brief, the following basic parameters are provided for a tracked CME: 
first appearance time and end time in the COR1 FOV, central PA, angular width, median and maximum velocity, and acceleration.
The code was rewritten in a more efficient and robust way.
Enhanced tracking algorithm provides more reliable physical CME parameters.

\subsection{3D Reconstructions with the Polarization Ratio Method} \label{subsubsec:PR_cor1_a}
Coronagraph images can reveal only 2D projections of a CME onto the plane of sky (POS), the structure of the CME in 3D space is still not fully known. Here we imply a method usually called the polarization ratio technique \citep{moran_three-dimensional_2004,dere_three-dimensional_2005,lu_measure_2017} to derive the 3D structure of a CME.
Unlike the traditional geometric methods \citep{thernisien_modeling_2006, feng_3d_apj_2012}, which relies on at least two perspective observations, the polarization ratio technique offers the capability of retrieving the 3D shape of a CME from a single perspective observation.
The technique was proposed based on the well-known Thomson scattering theory, according to which, the relationship between the polarization degree of the white light scattered by coronal electrons and the positions of the electrons can be established, e.g., along a given line of sight, the lower the polarization degree of the scattered light, the farther away the electrons from the POS.
The polarization degree of the coronal white light can be calculated from coronagraph polarimetric measurements, and then the weighted average distance of the coronal electrons along  LOS can be estimated.

The coronagraph polarimetric measurements are a set of three images observed at three different polarizer orientations.
From the three polarimetric coronagraph images, we can calculate the polarized and total brightness ($pB$ and $tB$), and thus the polarization degree (the ratio between pB and tB). Note that, before deriving $pB$ and $tB$ from polarimetric coronagraph images, the pre-event background emissions which contain dust-scattered (F-corona) and stray light are first subtracted from individual polarizer images.
To reduce the noise, the synthesized $pB$ and $tB$ images were both smoothed with a median filter method ($2\times2$ filter box).

Since the $M_{tracking}$ received from tracking results is the difference of CME's relative time motion, 
we use the convex hull method to obtain the preliminary CME region.
Then, $M_{smooth}$ is the smoothed CME regions derived by smoothing the edges of the preliminary CME region with circle filters of various sizes. 
We subtract the minimum FOV of COR1-A from M to simulate the complete range of CME structures.
Finally, the 3D position of the CME is calculated within the pre-defined mask.
The calculated results are displayed in the HEEQ coordinate system,  
which sets its origin at the center of the Sun,  and Z-axis along the solar rotation axis, and the X-axis in the plane containing the Z-axis and Earth. 
Note that, the polarimetric reconstruction suffers from ambiguities arising from Thomson scattering \citep{dai_cme_classification_2013}, e.g., from the polarization degree measurements, we can derive the average distance of CME away from the POS, but we are unable to determine which side of the POS the CMEs are moving on. To remove such ambiguity, solar ultraviolet (UV), Extreme UV images or magnetic field data revealing the source region activities of the CME are often used.

\section{Results and Comparisions} \label{sec:Results}

\subsection{Experimental Configurations}
SegFormer model is implemented on Python 3.8.10 and PyTorch 1.9.0 \citep{paszke_pytorch_2019}.
With regard to the model performance and GPU memory consumption,
the MiT-B1 encoder of SegFormer is selected,
the input image size is resized to $512 \times 512 \times 1$,
and the batch size is set at 4.
We trained the models using Adam \citep{kingma_adam_2017} optimizer for 250 epochs
with an initial learning rate of 0.0001.
Following the CosineAnnealingWarmRestarts \citep{loshchilov_sgdr_2017} schedule with $T_0=30$ and $T_{mult}=2$,
the learning rate changes dynamically during training.
The loss function uses the cross-entropy loss and the random seed is 3407.
Additionally, data augmentation techniques are employed, such as rotation, flipping, etc.
Data augmentation enhances the robustness and generalizability of the CME segmentation model,
especially when noise, gaussian blur, brightness, and contrast changes are added.

\subsection{Evaluation of Our Segmentation Results}
\begin{figure}[ht!]
  \plotone{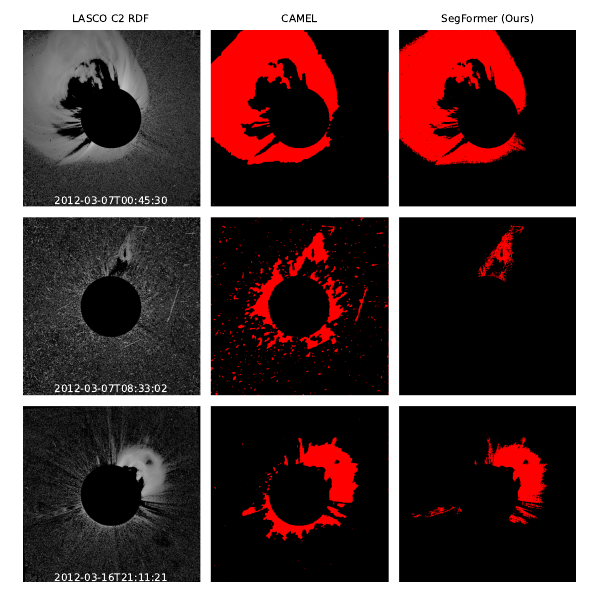}
  \caption{ 
    Our segmentation results vs. CAMEL on two CME events based on LASCO C2 images.
    From left to right: 
      LASCO C2 running-difference images, 
      CAMEL method on LASCO C2,
      our segmentation results on LASCO C2.
    From top to bottom are three CME events at different times.}
    \label{fig:segmentation}
\end{figure}
To evaluate our segmentation results, we use mean pixel accuracy (MPA) and mean intersection over union (MIoU). 
MPA calculates the proportion of pixels correctly classified for each class and then calculates the average across all categories. 
MIoU represents the result of the model summing and then averaging the ratio of the intersection of the predicted and true values for each class to their union. 
For this segmentation task, each pixel is classified into two classes: CME and non-CME. 
The evaluation metrics can be computed by the confusion matrix, in which higher values indicate better performance. 
Equation (\ref{equation:metrics}) shows the formula for these metrics.
\begin{equation} 
  \begin{aligned}
    MPA  &= \frac{1}{k+1} \sum_{i=0}^{k} \frac{ p_{ii} }{ \sum_{j=0}^k p_{ij} } \\
    MIoU &= \frac{1}{k+1} \sum_{i=0}^{k} \frac{ p_{ii} }{  \sum_{j=0}^{k} p_{ij} + \sum_{j=0}^{k} p_{ji}-p_{ii}} 
    \label{equation:metrics}
  \end{aligned}
\end{equation}
where $k$ represents the number of categories, $ k + 1$ represents the addition of the background class, 
$i$ is the true value from our segmentation dataset, $j$ is the predicted value, and $p_{ij}$ denotes the prediction of $i$ to $j$.
As seen in Table \ref{table:segmentation_compare}, 
the MIoU and MPA we achieved with the SegFormer model were 82.91\% and 98.05\%, which are significantly better than CAMEL \citep{wang_new_2019} with 48.58\% and 94.88\%, respectively.
Our evaluation showed that PSPNet \citep{zhao_pyramid_2017}, a CNN-based segmentation model, 
had inferior performance in terms of evaluation metrics compared to SegFormer.
As a result of the additional annotation of full halo CMEs, MIoU and MPA metrics are improved.
It should be noted that in the ablation experiment, we tried to delete random noise, brightness, and other data augmentation;
although both metrics were increased, the trained SegFormer model showed poor visualizations when tested on the data from COR1-A.
\begin{deluxetable*}{ccc}
  \tablenum{l} 
  \setlength{\tabcolsep}{7mm}
  \tablecaption{
    \label{table:segmentation_compare}
    Comparison of the segmentation accuracy
    }
  \tablewidth{0pt}
  \tablehead{
  \colhead{\textbf{Method}} 
  & \colhead{\textbf{MIoU}}
  & \colhead{\textbf{MPA}}
  }
  \startdata 
  CAMEL     & 48.58\%     & 94.88\% \\
  PSPNet    & 80.55\%     & 97.77\%  \\
  SegFormer & \textbf{82.91\%}     & \textbf{98.05\%} \\
  \enddata
\end{deluxetable*}

As shown in Figure \ref{fig:segmentation}, our segmentation results provide more finely-grained features than CAMEL.
The last two rows demonstrate that our method segments fewer abnormal noise points and focuses more on the CME structure. 
The panel shown in the segmentation module in Figure \ref{fig:flowchart} demonstrates that our segmentation model developed for LASCO C2 is also applicable to the COR1 coronagraph on aboard the STEREO-A satellite.
In terms of evaluation metrics and visualization results, our trained model SegFormer achieves a good balance.

\subsection{A One-year COR1-A 2D CME Catalog}
\begin{figure}[ht!]
  \plotone{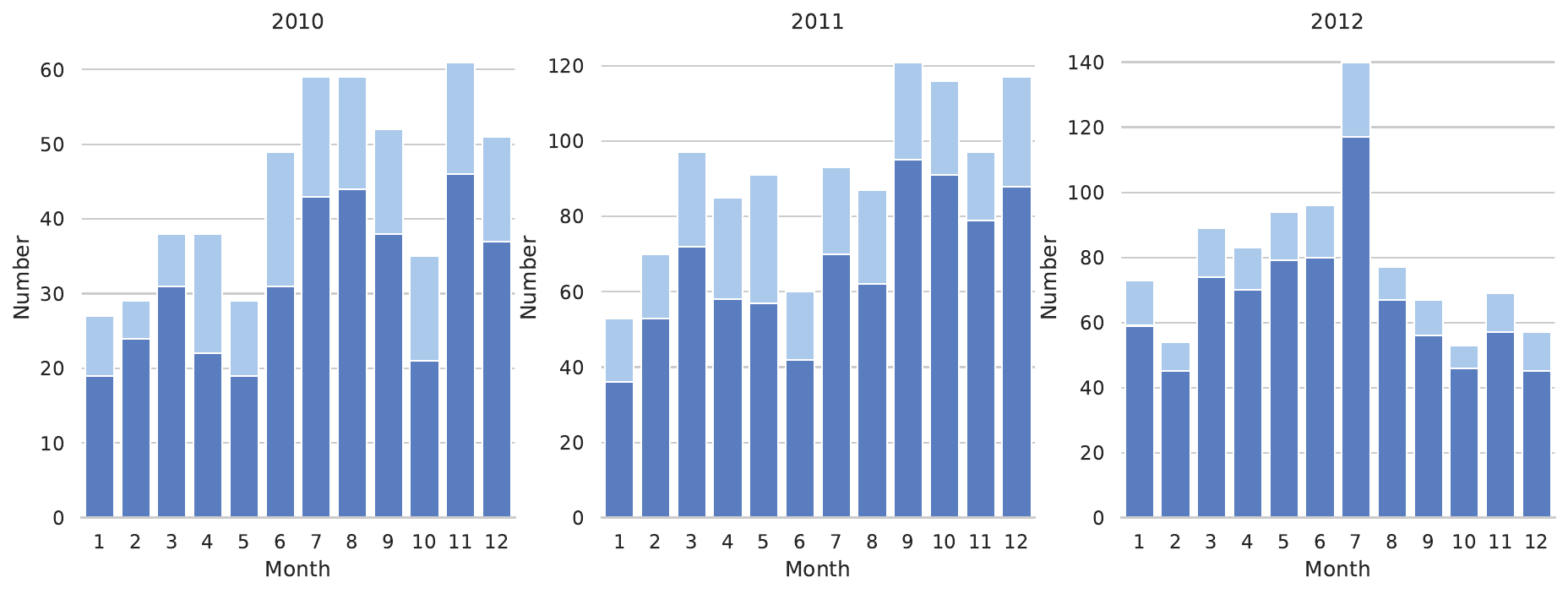}
  \caption{Each annual bar chart displays the number of CME events recorded per month from 2010 to 2012. Dark blue rectangular bars for each month indicate the number of events matched with the COR1-A Preliminary CME List, while light blue rectangular bars indicate unmatched CME events.}
\label{fig:tracking_cor1_three_years}
\end{figure}
\begin{figure}[ht!]
  \plotone{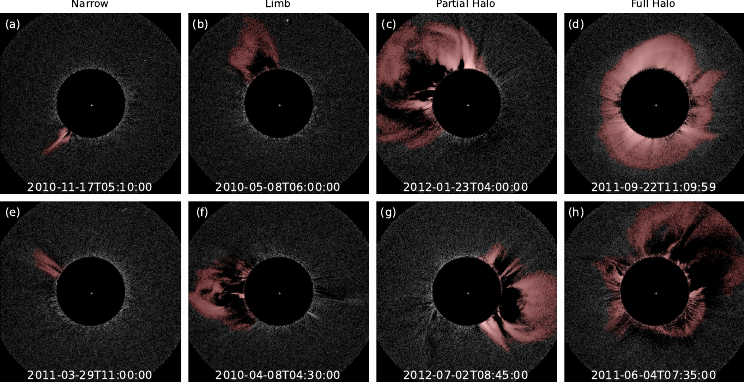}
  \caption{Tracking results for four different angular width distributions, with only one moment selected for each CME tracking result. Each column displays two CME events.
Panel (g) is available as an animation illustrating the temporal evolution of a partial halo CME event from our tracking results. This online animation runs from 08:05 UT to 09:05 UT on July 2 for COR1-A with a cadence of 5 minutes. The real-time duration of this animation is about 2 s.}\label{fig:cor1_four_angular_width}
    
\end{figure}

To test the generality of our method for detecting CMEs from LASCO C2 to COR1-A coronagraph images, 
we collected COR1-A observations for three years from 2010 to 2012.
Figure \ref{fig:tracking_cor1_three_years} shows the comparison between our tracking results and the preliminary list of CMEs provided 
by the COR1 manual catalog\footnote{https://cor1.gsfc.nasa.gov/catalog}.
In the Images/Event column, the first and last CME images recorded are used as the start and end times of the CME event.
It is inaccurate to assume that two CME events are the same CME simply 
by examining the start and end times of two CME events intersect.
It is possible that there are two wide CMEs propagating oppositely.
Therefore, the direction descriptions from the comments column were transformed into exact values, 
e.g., 90 degrees for E, 135 degrees for SE, and 120 degrees for ESE.
Since this manual catalog is only a preliminary abbreviated record without an accurate central PA, there may be some variance in central PAs compared to our results. 
For better matching, we removed the events that were described ``hardly/not seen/see in A (COR1-A)'' or did not mention the direction in COR1-A comments of the COR1 manual catalog.

CME events are more frequent in 2011 and 2012 when the Sun is more active \citep{2003ESASP.535..403G}, as illustrated in Figure \ref{fig:tracking_cor1_three_years}.
Based on our method, the match rates for three years were 
71\%, 74\%, and 83\%, respectively. 
Among them, the matching rate in 2012 was relatively high. 
In addition to the excellent performance of the segmentation model, we have also improved the tracking algorithm, 
especially solving the difficulty in distinguishing multiple CMEs from an image sequence,
which has contributed to the increase in matching rates.
Despite the fact that our method detected about 1.5 times more CME events than the COR1 Preliminary CME List and there were false positives, 
we also identified some CME events that were missed by this manual catalog.
In addition, we calculated that the difference in the average start time between the COR1 manual catalog 
and our catalog is within twenty minutes.

Due to the lack of an automated detection catalog for COR1-A and the availability of SEEDs\footnote{http://spaceweather.gmu.edu/seeds}  data limited to the COR2 and LASCO C2 catalogs, it is not feasible for us to make a comparison. 
As illustrated in Figure \ref{fig:cor1_four_angular_width}, we selected 8 CME events from the tracking results spanning three years, from 2010 to 2012, to showcase the identification results for four different types of CMEs with respect to their angular widths.

\subsection{A One-year COR1-A 3D CME Catalog}
Using the polarization ratio technique, we can determine the propagation direction of the CME, which can be used as input for forecasting CME space weather.
To evaluate the effectiveness of our automatic method, we compare our results with those calculated by \cite{mierla_3d_2009}, who applied the PR technique to the same data set COR1. 
In the Heliocentric Earth Equatorial (HEEQ) coordinate system, we estimated the average longitude, latitude, and altitude of three identical CME events.
For the background subtraction, we tried two types: 1-day minimum background and the pre-event background.
As shown in Table \ref{table:reconstruction}, our results obtained by subtracting the pre-event images are more consistent with those calculated by \cite{mierla_3d_2009}.
The differences are supposed to originate from the selection of different CME regions, the subtraction of different backgrounds, etc.
With the STEREO-A EUVI observations of these three CMEs, the CME source regions are all located on the foreside of the Sun.
\begin{deluxetable*}{ccccll}
  \tablenum{2} 
  \tablecaption{
    \label{table:reconstruction}
    Error analysis for reconstruction result.
    The parameters of 3D CME reconstruction are compared with another PR method at three different times.
    The COR1 data were recorded at 19:15\,UT on 15 May 2007, 21:30\,UT on 31 August 2007, and 19:05\,UT on 25 March 2008. 
    Referring to the HEEQ coordinate system, the mean values of the parameters for the points reconstructed using different methods applied to COR1 data: 
    fore-side longitude, fore-side latitude, 
    and the center distance of the CME from the Sun center.
    }
  \tablewidth{0pt}
  \tablehead{
  \colhead{Time} & \colhead{Method} 
  & \multicolumn{2}{c}{\makecell[c]{Foreside Propagation \\ \hline Longitude \quad Latitude}}
  & \colhead{Rcenter} 
  & \colhead{Dots Num} 
  }
  \startdata 
  {}                  & $PR_\text{{pre-event}}$              & -55.0   & -13.3    & 1.89 & 1881 \\
  2008-03-25T19:05:00 & $PR_\text{{day-minimum}}$            & -42.3   & -14.7   & 2.02 & 9169 \\
  {}                  & Mierla et al. 2009     & -58      & -12      & 1.8  & -   \\  
  \hline  
  {}                  & $PR_\text{{pre-event}}$              & 79.8    & -23.4   & 2.13 & 6447 \\
  2007-08-31T21:30:00 & $PR_\text{{day-minimum}}$            & 67.3    & -19.2   & 2.45 & 12520  \\
  {}                  & Mierla et al. 2009     & 87       & -31      & 1.96  & -   \\
  \hline
  {}                  & $PR_\text{{pre-event}}$              & -62.9   & 12.7    & 2.11 & 4878 \\
  2007-05-15T19:15:00 & $PR_\text{{day-minimum}}$            & -48.5   & 13.0    & 2.41 & 14165  \\
  {}                  & Mierla et al. 2009     & -72      & 8        & 2.0  & -   \\  
  \enddata
\end{deluxetable*}

\begin{figure}[ht!]
    \plotone{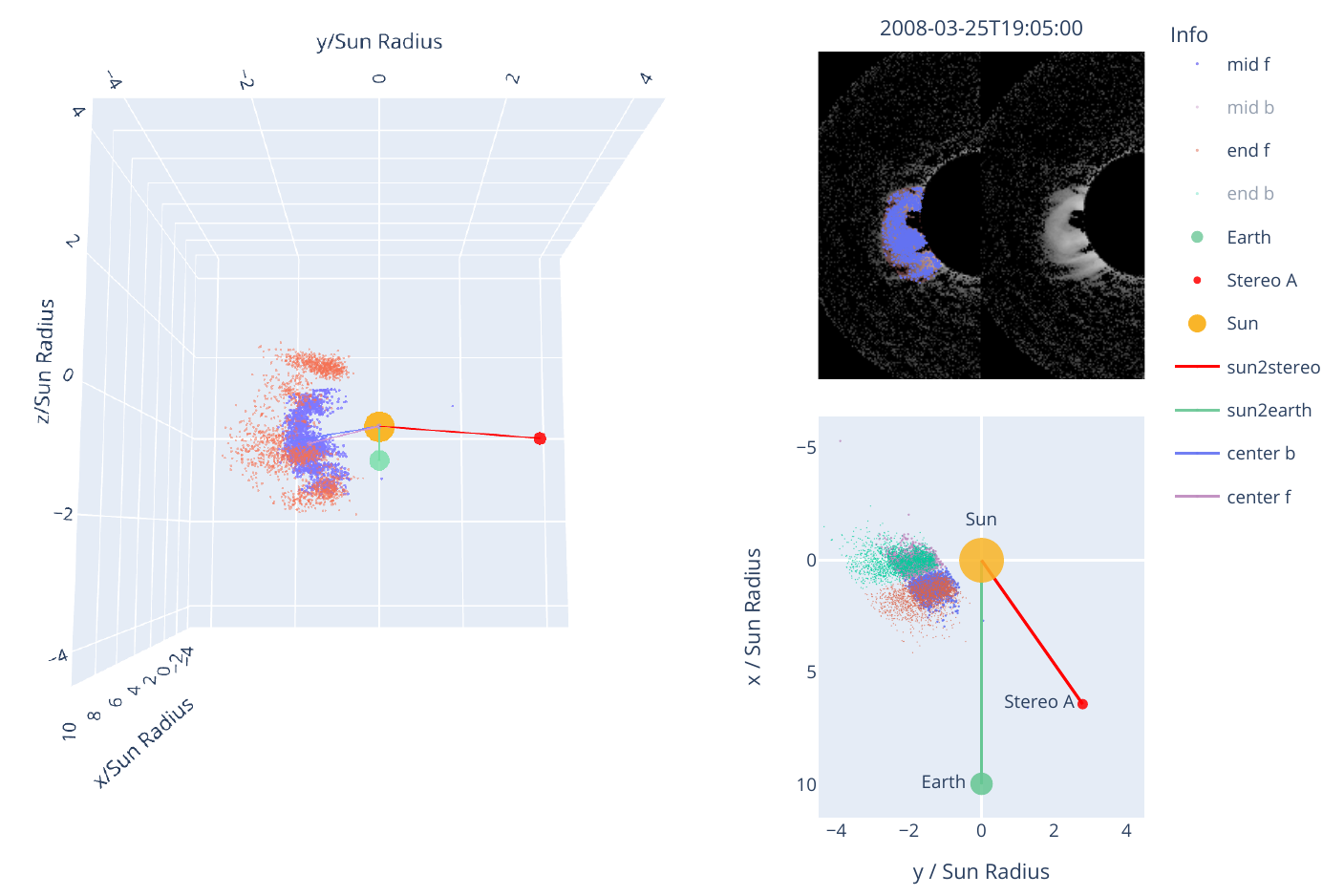}
    \caption{Multi-viewpoint visualization of CMEs in 3D space.
      The left panel shows the positional relationship between the points of 3D CME reconstruction on the foreside and the sun, the earth, and STEREO-A. 
      Red dots and blue dots indicate the CME evolution distribution at 19:05 and 19:35 on March 25, 2008, respectively.
      In the right top panel, blue dots represent 3D reconstructed points projected onto STEREO-A's total brightness image. 
      The right bottom panel is a top view of the left panel and the positions of the CME at two time moments on the other side are added.\label{fig:multi_3d_cme}}
\end{figure}

Figure \ref{fig:multi_3d_cme} shows a multi-viewpoint visualization of the CME at two moments in time in 3D space.
To better visualize the distribution of CMEs on the left panel, 
we chose to hide the backside CME.
By clicking the legends in the visualization, we can selectively hide or show the parts we are interested in.
It is the first catalog\footnote{https://github.com/h1astro/CAMEL-II} to provide the automatic 3D spatial distribution of CMEs from a single view. We provided 3D reconstruction results at two moments, including average distance, 3D velocity, foreside longitude and latitude, backside longitude and latitude, etc. For a CME event, its 3D geometric centers at different moments can be computed from the 3D CME coordinates. Subsequently, the 3D propagation direction of the CME can be obtained from the geometric centers.

\section{Summary and Conclusion} \label{sec:summary}
In this paper, we have developed a novel automatic system for the classification, segmentation, tracking, and 3D reconstruction of CMEs with STEREO/COR1-A coronagraph images.
Using LASCO C2 observations and records from the CDAW catalog, we annotated a semantic segmentation dataset containing four different types of CMEs with respect to their angular widths, i.e., narrow, limb, partial halo and full halo CMEs.
The Transformer-based segmentation model was utilized to segment CMEs for the first time.
The SegFormer model provides more finely-grained features compared to CAMEL and exhibits better robustness and generalizability in the visualization of the images based on COR1-A.
SegFormer outperforms CAMEL by 34.33\% in MIoU and 3.14\% in MPA, which indicates that SegFormer achieves state-of-the-art performance while maintaining a good balance between evaluation metrics and visualization outcomes.
One advantage of Transformer-based models over CNNs is their larger effective receptive field, 
which facilitates global reasoning.

We published the first COR1-A one-year catalog of CME events.
To track CMEs in various segmented image sequences as accurately as possible, our tracking algorithm has undergone a lot of modifications.
In particular, we solved the difficulty in distinguishing multiple CMEs from an image sequence.
The improved tracking algorithm decreases the rate of missed detection and false detection and provides more reliable CME physical parameters.

Currently, no publicly available catalog provides 3D structures of CMEs based on single-view observations. 
The main reason for this is the labor-intensive and time-consuming nature of manually masking out CME structure areas. 
To overcome this limitation, we propose an automated approach. 
Instead of relying on manual masking, we utilize filter operations of the tracking results to obtain mask areas. 
Subsequently, we employ polarization ratio techniques to determine the center of mass along the line of sight and estimate the density distribution of the CME. 
This automated approach streamlines the process and eliminates the need for manual masking, allowing for a more efficient analysis of CME structures.

New approaches proposed in this work can be summarised as follows:
i) Introducing a new system automating the CME detection and 3D reconstruction.
ii) Investigating the performance of Transformer-based networks on segmenting CMEs. 
iii) Proposing a method for automated 3D reconstruction of CMEs.
iv) Publishing our CME semantic segmentation dataset to improve the application of deep learning in CME beyond image classification.
v) Publishing an automated catalog of CME events with a high detection rate.
vi) Validating the generalization of this system from LASCO observations to COR1-A observations. 
In the future, our technique can be migrated to any white-light coronagraph data with polarization measurements.

The tracking of CMEs is complex, particularly during periods of high solar activity when multiple CMEs occur closely in time or even simultaneously.
Our tracking algorithms need further refinement to distinguish multiple CMEs in more difficult cases with intersecting PAs but differing occurrence times in an image sequence.
The combination of automatic recognition of the CME source region as indicated by flares, waves, eruptive filaments, etc., can aid in determining the foreside or backside uncertainty of the 3D CME inherited in the Thomson Scattering.

\section*{}
STEREO is a project of NASA, SOHO a joint ESA/NASA project.
The SECCHI data used here were produced by an international consortium of the NRL, LMSAL, NASA GSFC (USA), RAL and Univ. Birmingham (UK), MPS (Germany), CSL (Belgium), IOTA and IAS (France). 
This work is supported by the Strategic Priority Research Program of the Chinese Academy of Sciences.
Grant No. XDB0560000, National Key R\&D Program of China 2022YFF0503003 (2022YFF0503000), NSFC (grant Nos. 11973012, 11921003, 12103090,12203102), the mobility program (M-0068) of the Sino-German Science Center. This work benefits from the discussions of the ISSI-BJ Team ``Solar eruptions: preparing for the next generation multi-waveband coronagraphs"

\bibliography{bibcme}{}
\bibliographystyle{aasjournal}

\end{document}